\documentclass[twocolumn,amsmath,amssymb,aps,pra,longbibliography,floatfix]{revtex4-1}
\usepackage{physics}
\usepackage{units}
\usepackage{amsmath}
\usepackage{cases}
\usepackage{url}
\usepackage{natbib}
\usepackage{textcase}
\usepackage{amssymb}
\usepackage{graphicx}
\usepackage{bm} 

\usepackage{times}
\usepackage{float}
\usepackage{multirow,microtype,color,relsize,ulem}
\usepackage{subfigure}
\usepackage[breaklinks=true]{hyperref}
\usepackage[utf8]{inputenc}
\usepackage[english]{babel}
\hypersetup{colorlinks=true,linkcolor=blue,citecolor=blue}
\hypersetup{linktocpage}
\usepackage{CJK}
\usepackage{breakcites}
\usepackage{float}
\usepackage[dvipsnames]{xcolor}
\definecolor{mypine}{RGB}{1, 121, 111}

\begin{document}
\begin{CJK*}{UTF8}{gbsn}
\title{Biexciton crystal in a two-dimensional semiconductor heteropentalayer.}

\author{Yi Huang~(黄奕)}
\email[Corresponding author: ]{huan1756@umn.edu}
\author{B.\,I. Shklovskii} 
\affiliation{School of Physics and Astronomy, University of Minnesota, Minneapolis, Minnesota 55455, USA}

\received{\today}

\begin{abstract}
This paper is written for the Special Issue in Honor of Emmanuel Rashba.
We study the gas of indirect dipolar excitons created by an interband illumination of a pentalayer WSe$_2$/MoSe$_2$/WSe$_2$/MoSe$_2$/WSe$_2$. We show that two colinear indirect excitons bind into a linear biexciton with twice larger dipole moment. Two biexcitons with opposite dipole directions attract each other at large distances and repel each other at short distances. Therefore, biexcitons form a staggered crystal with anti-ferroelectric square lattice. The electrostatic energy of this crystal per biexciton has a minimum at the biexciton concentration $n =n_c=0.14d^{-2}$, where $d \sim 0.7$ nm is a single layer thickness. At small illumination intensity, biexcitons condense into sparse crystallites with $n = n_c$, where photoluminescence frequency is red shifted and independent on the light intensity. 
We also study a capacitor made of five identical semiconductor monolayers separated by hBN spacers where a critical voltage applied between layers 1, 3, 5, and 2, 4 abruptly creates a similar biexciton crystal. At this voltage, the differential capacitance diverges.
\end{abstract}

\date{\today}

\maketitle
\end{CJK*}
\section{Introduction}\label{sec:intro}

\begin{figure}[t]
	\includegraphics[width=0.5\linewidth]{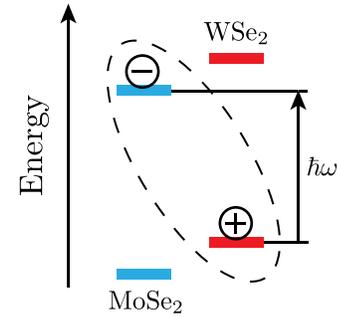}
	\caption{Band alignment in MoSe$_2$/WSe$_2$ bilayer. Absorption of a light quantum $\hbar\omega$ results in an electron in MoSe$_2$ layer and a hole in WSe$_2$ layer bound by the Coulomb interaction in an spatially indirect exciton.} \label{fig:optical}
\end{figure}

Heterostructures of two-dimensional (2D) transition metal dichalcogenides (TMDC) demonstrated wide spectrum of correlated states of excitons. 
For example~\cite{Rivera2015,Calman2020}, in a bilayer MoSe$_2$/WSe$_2$, the type II band alignment of MoSe$_2$ and WSe$_2$ monolayers allows the formation of spatially indirect excitons, in which an electron in MoSe$_2$ binds to a hole in WSe$_2$ (see Fig.\ref{fig:optical}). Parallel dipolar excitons repel each other at all distances between them or at all two dimensional exciton concentrations $n$ tuned by the illumination intensity.  
Because of the weak overlap of the electron and hole wave functions, these excitons decay slowly enough to form the correlated ground state, which minimizes their repulsion. It was observed~\cite{Rivera2015,Calman2020} that in MoSe$_2$/WSe$_2$ devices, the indirect exciton luminescence line is blue shifted due to the dipole-dipole repulsion, and this effect becomes stronger as the illumination intensity increases. 

\begin{figure}[t]
	\includegraphics[width=\linewidth]{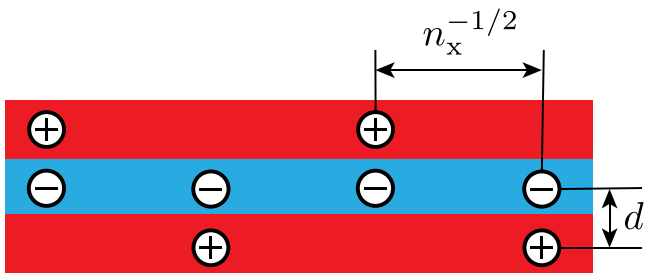}	
	\caption{Schematic of a trilayer WSe$_2$/MoSe$_2$/WSe$_2$. WSe$_2$ (first and third) layers are red and MoSe$_2$ (second) layer is blue. When the device is illuminated at low temperatures, the type II band alignment of neighboring WSe$_2$/MoSe$_2$ heterostructures allows the formation of indirect excitons consisting of an electron in MoSe$_2$ and a hole in WSe$_2$. Excitons of opposite polarity attract each other and form a staggered crystal with alternating exciton dipoles.}
\label{fig:trilayer}
\end{figure}

\begin{figure}[t]
	\includegraphics[width=0.5\linewidth]{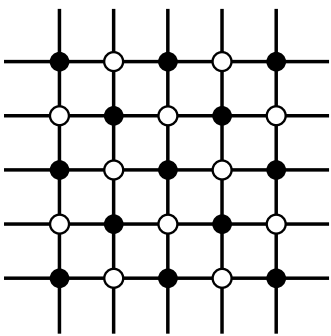}
	\caption
	{The top view of an anti-ferroelectric square lattice of alternating dipolar excitons or biexcitons. Black (white) circles correspond to dipoles whose orientation points up (down). Straight lines are guides for eye.} \label{fig:squarelattice}
\end{figure}

It was predicted recently~\cite{Sammon2019} that in a trilayer device WSe$_2$/MoSe$_2$/WSe$_2$ (WMW)~\cite{Plochocka2017,Choi2018} and similar devices with identical hexagonal boron nitride (hBN) spacers between layers, repulsion of excitons at small distances is replaced by their attraction at larger distances. 
This happens because in a WMW device, the hole of an exciton originates with equal probability in either the upper or lower WSe$_2$ layers, while the electron is always at the middle layer (see Fig.~\ref{fig:trilayer}). At large distances, two oppositely oriented dipolar excitons attract each other, while the repulsion takes over at short distances, because electrons located in the middle layer approach each other. 
It was shown that the interaction energy of directed up and down dipolar excitons has a minimum near $n_{\mathrm x}^{-1/2} = 2.8d$, where $n_{\mathrm x}$ is the two-dimensional concentration of excitons and $d\sim 0.7$ nm~\cite{Slobodkin2020} is the distance between two layers. 
The ground state of these excitons in a WMW device is a staggered square lattice of alternating up and down dipoles (see Fig.~\ref{fig:squarelattice}). 
The interaction energy per exciton of such a crystal has a minimum at the concentration $n_{{\mathrm x}c} = 0.12d^{-2}$. 
Therefore, at low illumination intensities, when $n_{\mathrm x} \ll n_{{\mathrm x}c}$, all excitons condense into droplets of density $n_{{\mathrm x}c}$, which do not interact with each other. 

Ref.~[\onlinecite{Sammon2019}] ignored the tunneling of an exciton hole between two possible locations in the lower and upper WSe$_2$ layer. 
For a single exciton this tunneling hybridizes two opposite orientations of a dipole moment and creates a quadrupolar dark exciton~\cite{Slobodkin2020}. However, in a system of many interacting excitons this happens only at small enough concentration $n$ when the energy splitting between symmetric and anti-symmetric hole states is larger than the electrostatic energy required to invert the orientation of an exciton dipole in the staggered classical crystal. According to Ref.~[\onlinecite{Slobodkin2020}] in WMW devices, the quadrupolar exciton phase marginally loses to the staggered crystal at $n_{\mathrm x} \sim n_{{\mathrm x}c}$.

A recent experimental paper~\cite{Bai2022} used optical spectroscopy to study exciton phases in a WMW device gated by two transparent graphene layers separated by $\sim 10$ nm hBN spacers. The authors discovered the crystalline phase of indirect excitons. Indeed, they observed sharper peaks in photoluminescence due to stronger localization of excitons.
They also measured directly the diffusion coefficient of excitons and showed that it decreases in the crystalline phase. The authors attributed the low density dark exciton state to the predicted quadrupolar exciton~\cite{Slobodkin2020}, and showed how the transition from dark to bright exciton crystal happens with increasing exciton concentration or with growing electric field. 

\begin{figure}[t]
	\includegraphics[width=\linewidth]{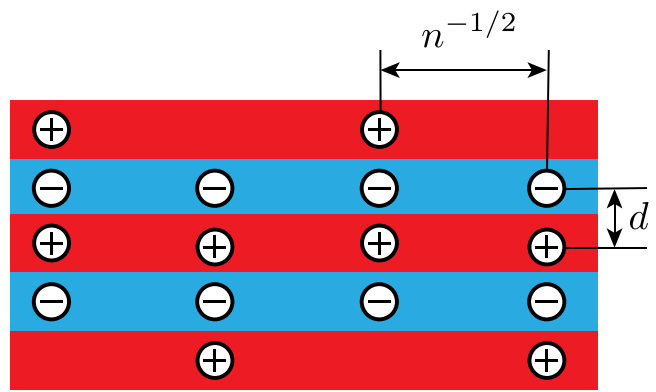}
	\caption{Pentalayer WSe$_2$/MoSe$_2$/WSe$_2$/MoSe$_2$/WSe$_2$ device. WSe$_2$ layers are red and MoSe$_2$ layers are blue.  When the device is illuminated at low temperatures, linear biexcitons with alternating orientations are created. Biexcitons form a staggered square lattice, and the top view of which is shown in Fig.\,\ref{fig:squarelattice}.}
	 \label{fig:pentalayer}
\end{figure}

In this paper we turn to TMDC pentalayers  WSe$_2$/MoSe$_2$/WSe$_2$/MoSe$_2$WSe$_2$ (WMWMW). 
We claim that in a pentalayer, as shown in Fig.~\ref{fig:pentalayer}, spatially indirect excitons of two consecutive bilayers attract each other and form linear biexcitons~\footnote{Similar exciton attraction was previously studied for two close bilayer systems  ~\cite{Cohen2016,Hubert2019,Hubert2020,Choksy2021}}.
Two biexcitons with opposite directions of their dipole moments attract each other at large distances in the plane of the pentalayer. On the other hand, at small distances they repel each other because they have in-plane charges of the same sign in the three middle layers. This competition of biexciton  attraction and repulsion leads to an anti-ferroelectric biexciton crystal in a WMWMW pentalayer. The top and side view of the biexciton crystal are shown in Fig.~\ref{fig:squarelattice} and \ref{fig:pentalayer}.

\begin{figure}
 	\includegraphics[width=\linewidth]{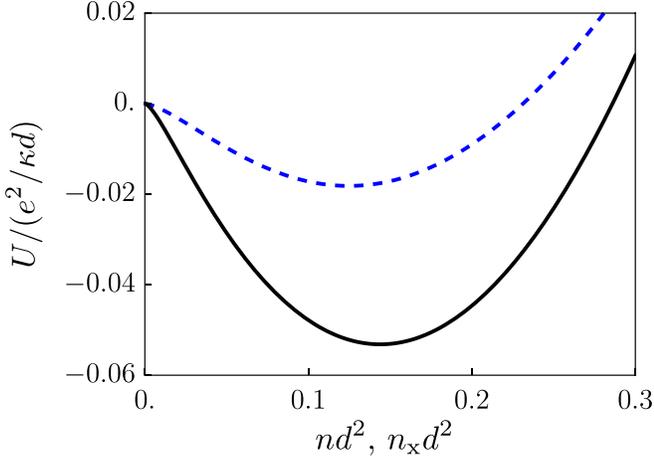}
 	\caption{The full black curve shows the dimensionless electrostatic energy per biexciton $U/(e^2/\kappa d)$ in a staggered lattice (c.f. Fig.~\ref{fig:squarelattice}) versus the dimensionless biexciton density $nd^2$ in a WMWMW device. The blue dashed curve reproduces the electrostatic energy of a staggered lattice per exciton versus the dimensionless density of excitons $n_{\mathrm x}d^2$ in a WMW device~\cite{Sammon2019}.}
 \label{fig:energy}
 \end{figure}

The electrostatic energy of the staggered crystal per biexciton in a pentalayer WMWMW is calculated below. 
The result is shown by the full black curve as a function of $nd^2$ in Fig.~\ref{fig:energy}, where $n$ is the 2D concentration of biexcitons. This curve has a minimum at $n=n_c=0.14d^{-2}$. 
In the same figure, by the blue dash curve we reproduce the electrostatic energy per exciton of a staggered crystal in a WMW trilayer as a function of $n_{\mathrm x}d^2$, where $n_{\mathrm x}$ is the concentration of excitons~\cite{Sammon2019}. 
We see that although both energies have minimums at close values of $nd^2$ or $n_{\mathrm x}d^2$, the minimum of cohesive energy of biexcitons is approximately 3 times deeper. This allows to study the biexciton crystal in a WMWMW device at higher temperatures. 
We also show below that, in the pentalayer WMWMW, hybridization of biexcitons with opposite directions of their dipole moments into dark quadrupolar boexcitons plays a minor role, so that the staggered anti-ferroelectric crystal of biexcitons dominates the phase diagram. 

We therefore predict that at $n < n_c$ biexcitons condense into crystallites with $n = n_c$. This crystallization can be observed spectroscopically, via additional to isolated biexciton red shift of the exciton photoluminescence. 
It originates due to interaction of recombining an electron-hole pair with surrounding biexciton crystal, grows during condensation, and saturates at the level independent on the illumination intensity, when the final density of the biexciton crystal $n_c$ is reached.

The plan of the rest of the paper is as follows. 
In section~\ref{sec:comp}, we address the issue of biexciton hybridization and show that in the pentalayer case it can be neglected. 
In section~\ref{sec:energy}, we present our calculation of the interaction energy of the biexciton crystal.
In section~\ref{sec:flip}, we calculate of the energy cost to flip the orientation of a biexciton in a fixed staggered crystal.
In section~\ref{sec:capacitor}, we consider a device made of five layers of MoSe$_2$ separated by identical hBN spacers and show that a critical value of voltage applied to its layers 1, 3, 5 with respect to layers 2, 4 abruptly induces a finite-concentration biexciton crystal. 
This means that at this voltage, the pentalayer 135-24 differential capacitance is infinite.
In section~\ref{sec:fourlayer}, we briefly comment on the four-layer contactless device.
 
\section{Competition between dipolar and quadrupolar biexcitons }
\label{sec:comp}

Let us show that the staggered crystal of dipolar biexcitons in a WMWMW device easily survives the competition of a crystal of quadrupolar dark biexcitons. 
Indeed, there is a possibility that the quantum mechanical tunneling of holes between bottom and top WSe$_2$ layers hybridizes two biexcitons with opposite orientations of their dipole moments into a symmetric quadrupolar dark biexciton. 
Whether the quadrupolar biexcitons win depends on the relationship between the quantum energy splitting of symmetric and antisymmetric biexciton states and the classical electrostatic energy cost of inverting the direction of a single biexciton in a fixed staggered lattice of its neighbors. 
We know that near $n_{\mathrm x}=n_{{\mathrm x}c}$ in a WMW device, the staggered biexciton crystal phase has smaller energy and marginally dominates the phase diagram~\cite{Slobodkin2020}. Now we should look at this competition in a pentalayer device. 

\begin{figure}[t]
 	\includegraphics[width=\linewidth]{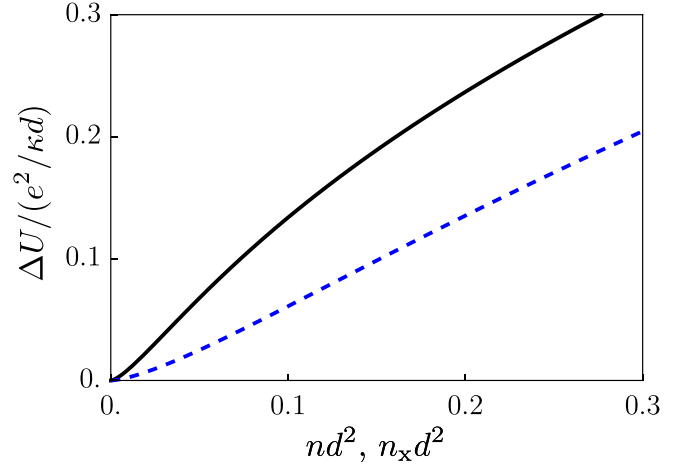}
 	\caption{The full black curve shows the dimensionless energy cost $\Delta U/(e^2/\kappa d)$ of inverting of a biexciton dipole moment in a pentalayer device (c.f. Fig.~\ref{fig:pentalayer}) by moving a hole between the top WSe$_2$ layer and the bottom WSe$_2$ layer versus the dimensionless biexciton density $nd^2$. The blue dashed curve shows the energy of inverting an exciton in a trilayer device (c.f. Fig.~\ref{fig:trilayer}) by moving a hole from the top WSe$_2$ layer to the bottom one versus the exciton density $n_{\mathrm x}d^2$.}
 \label{fig:cost}
 \end{figure}

Below we calculate the Coulomb energy cost $\Delta U$ of inverting the orientation of a dipolar biexciton in a WMWMW device. 
It is plotted as a function of $nd^2$ in Fig.~\ref{fig:cost} together with the cost of inverting a dipolar exciton for a WMW device as a function of $n_{\mathrm x}d^2$. 
We see that the former is approximately twice larger than the latter at small $n$ and $n_{\mathrm x}$.
On the other hand, the quantum energy splitting in a WMWMW device is much smaller than in a WMW device. Indeed, the tunneling of a hole from the top WSe$_2$ layer to the bottom one requires simultaneous hopping of two holes. The hole from the middle WSe$_2$ layer hops to the bottom WSe$_2$ layer, while another hole of the same biexciton hops from the top WSe$_2$ layer to the middle WSe$_2$ layer. 
As a result in a WMWMW device, the quantum energy splitting looses the competition by a large margin and our classical theory is applicable practically at all interesting concentrations.  
Together with a more stable staggered lattice due to a larger interaction energy per biexciton, this makes the physics of a WMWMW device simpler than that of a WMW one.

\section{Calculation of the energy of staggered biexciton crystal}
\label{sec:energy}
In the next two sections, we explain how we arrived at results shown in Fig.~\ref{fig:energy} and Fig.~\ref{fig:cost}.
Without loss of generality, we pick in Fig.~\ref{fig:pentalayer} a biexciton with its dipole moment pointing up as the reference point, and calculate its interaction with all other charges of the lattice. For convenience we separate $U$ into four terms $U = U_{h1} + U_{e1} + U_{h2} + U_{e2}$, where $h1,\,h2$  ($e1,\,e2$) label the first and second holes (electrons) counting from the top in the reference biexciton. In the calculation of the biexciton interaction energy we ignore the interaction between electrons and holes of the same biexciton which are responsible only for its self-energy.

Each of four terms of $U$ consists of contributions of two sublattices labeled by $\bullet$ (up) and $\circ$ (down) in Fig.~\ref{fig:squarelattice} ($r_{\alpha}$ is the 2D coordinate of a biexciton $\alpha$):
\begin{align}
    U_{h1} = \frac{e^2}{2\kappa} \sum\limits_{\alpha }^{\circ} &\left(\frac{1}{\sqrt{r_{\alpha}^2 + (2d)^2}} + \frac{1}{\sqrt{r_{\alpha}^2 + (4d)^2}} \right.\nonumber\\
    &- \left.\frac{1}{\sqrt{r_{\alpha}^2 + d^2}} - \frac{1}{\sqrt{r_{\alpha}^2 + (3d)^2}}\right) \nonumber\\
    +\frac{e^2}{2\kappa} \sum\limits_{\alpha \neq 0}^{\bullet} &\left(\frac{1}{\sqrt{r_{\alpha}^2}} + \frac{1}{\sqrt{r_{\alpha}^2 + (2d)^2}} \right.\nonumber\\
    &- \left.\frac{1}{\sqrt{r_{\alpha}^2 + d^2}} - \frac{1}{\sqrt{r_{\alpha}^2 + (3d)^2}}\right),
\end{align}
\begin{align}
    U_{e1} = \frac{e^2}{2\kappa} \sum\limits_{\alpha }^{\circ} &\left(\frac{1}{\sqrt{r_{\alpha}^2 }} + \frac{1}{\sqrt{r_{\alpha}^2 + (2d)^2}}  \right.\nonumber\\
    &- \left.\frac{1}{\sqrt{r_{\alpha}^2 + d^2}} - \frac{1}{\sqrt{r_{\alpha}^2 + (3d)^2}}\right) \nonumber\\
    +\frac{e^2}{2\kappa} \sum\limits_{\alpha \neq 0}^{\bullet} &\left(\frac{1}{\sqrt{r_{\alpha}^2}} + \frac{1}{\sqrt{r_{\alpha}^2 + (2d)^2}} - \frac{2}{\sqrt{r_{\alpha}^2 + d^2}} \right),
\end{align}
\begin{align}
    U_{h2} = \frac{e^2}{2\kappa} \sum\limits_{\alpha \neq 0} \left(\frac{1}{\sqrt{r_{\alpha}^2}} + \frac{1}{\sqrt{r_{\alpha}^2 + (2d)^2}} - \frac{2}{\sqrt{r_{\alpha}^2 + d^2}} \right),
\end{align}
\begin{align}
    U_{e2} = \frac{e^2}{2\kappa} \sum\limits_{\alpha}^{\circ} &\left(\frac{1}{\sqrt{r_{\alpha}^2}} + \frac{1}{\sqrt{r_{\alpha}^2 + (2d)^2}} - \frac{2}{\sqrt{r_{\alpha}^2 + d^2}} \right) \nonumber\\
    +\frac{e^2}{2\kappa} \sum\limits_{\alpha  \neq 0}^{\bullet} &\left(\frac{1}{\sqrt{r_{\alpha}^2 }} + \frac{1}{\sqrt{r_{\alpha}^2 + (2d)^2}}  \right.\nonumber\\
    &- \left.\frac{1}{\sqrt{r_{\alpha}^2 + d^2}} - \frac{1}{\sqrt{r_{\alpha}^2 + (3d)^2}}\right),
\end{align}
where $\kappa$ is the dielectric constant of the system and $\alpha \neq 0$ means we exclude the coordinate origin (a dipole pointing up) in the summation.
As a result, we obtains the total interaction energy 
\begin{align}\label{eq:U_total}
    U = \frac{e^2}{\kappa d} f(nd^2),
\end{align}
where the dimensionless function $f(x)$ is given by
\begin{gather}
    f(x) = 2 \sqrt{x} \sum\limits_{i = 1}^{\infty} \sum\limits_{j = 0}^{\infty} \left(\frac{4}{\sqrt{i^2 + j^2 }}
    + \frac{4}{\sqrt{i^2 + j^2 + 2^2 x}} \right.\nonumber \\
    - \left.\frac{6}{\sqrt{i^2 + j^2 + x}} -\frac{2}{\sqrt{i^2 + j^2 + 3^2 x}} \right) \nonumber \\
    +2 \sqrt{x} \sum\limits_{i = 0}^{\infty} \sum\limits_{j = 0}^{\infty}\qty(2 - \delta_{j0} )\left(\frac{1}{\sqrt{(2i + 1)^2 + (2j)^2 + 4^2x}} \right. \nonumber \\
    - \left.\frac{1}{\sqrt{(2i + 1)^2 + (2j)^2}}\right).
\end{gather}
The result is shown by the full black curve as a function of $nd^2$ in Fig.~\ref{fig:energy}, where $n$ is the 2D concentration of biexcitons.

Comparison with WMW trilayer ~\cite{Sammon2019} shows that minimum of energy is 3 times deeper in a pentalayer. The reason for this is that biexcitons in a pentalayer device have twice larger dipole moment than excitons in a trilayer device. Thus, the absolute value of the attraction energy is four times larger. Repulsion between two biexcitons at small distances is substantially larger than the case of excitons as well. This happens because for biexcitons it originates from same-sign charges repulsion in three layers instead of one layer charge repulsion for excitons in the WMW device. 

\section{Calculation of the energy cost of biexciton dipole moment inversion}
\label{sec:flip}
Next we calculate the Coulomb energy cost to flip the orientation of a biexciton in the staggered lattice of a WMWMW pentalayer (c.f. Fig.~\ref{fig:pentalayer}).
By moving the hole $h1$ of a reference biexciton in the top WSe$_2$ layer to the bottom WSe$_2$ layer, i.e. to the position which we label as $h3$, we find its electrostatic energy
\begin{align}
    U_{h3} = \frac{e^2}{2\kappa} \sum\limits_{\alpha }^{\circ} &\left(\frac{1}{\sqrt{r_{\alpha}^2}} + \frac{1}{\sqrt{r_{\alpha}^2 + (2d)^2}} \right.\nonumber\\
    &- \left.\frac{1}{\sqrt{r_{\alpha}^2 + d^2}} - \frac{1}{\sqrt{r_{\alpha}^2 + (3d)^2}}\right)\nonumber\\
    +\frac{e^2}{2\kappa} \sum\limits_{\alpha \neq 0}^{\bullet} &\left(\frac{1}{\sqrt{r_{\alpha}^2 + (2d)^2}} + \frac{1}{\sqrt{r_{\alpha}^2 + (4d)^2}} \right.\nonumber\\
    &- \left.\frac{1}{\sqrt{r_{\alpha}^2 + d^2}} - \frac{1}{\sqrt{r_{\alpha}^2 + (3d)^2}}\right).
\end{align}
The energy cost of inverting a biexciton orientation is then given by $\Delta U = U_{h3} - U_{h1}$. 
This energy is plotted as the full black curve in Fig.~\ref{fig:cost}, where the energy cost to flip an exciton for a WMW trilayer is shown by the dashed blue curve for comparison.
At very small concentrations $nd^2,n_{\mathrm x}d^2 \to 0$, we have $\Delta U \propto n^{3/2},n_{\mathrm x}^{3/2}$ with the coefficient of a pentalayer device 4 times larger than that for a trilayer.
At larger concentration $n,n_{\mathrm x} \gtrsim n_c = 0.14 d^{-2}$, the energy cost to flip a biexciton in a pentalayer is 2 times larger than to flip an exciton in a trilayer. 

Above we used the three-dimensional Coulomb interaction between charges, because we assumed that all studied devices have relatively thick hBN layers below and above them which make dielectric constant contrast relatively small. 
These hBN layers are used for gating and are not shown in Figs.~\ref{fig:trilayer} and \ref{fig:pentalayer}.

Our theory also assumes that the electrons (holes) are point-like particles located in the middle of a layer. This is a good assumption if $n^{-1/2} \gg a$, where $a$ is the localization length of electrons (holes).
Using an estimate $a =\hbar/\sqrt{2m^{\star} E_b}$, where the effective mass $m^{\star} \sim 0.5 m_0$~\footnote{For electrons in MoSe$_2$ the effective mass is $m^{\star}_e = 0.49\,m_0$~\cite{Liu2020,Liu2021} and for holes in WSe$_2$ is $m^{\star}_h = 0.42\,m_0$~\cite{Nguyen2019}.}, $m_0$ is the free electron mass, and $E_b \sim 250$ meV~\cite{Gillen2018,Kamban2020} is the binding energy of an exciton, we get $a \sim 0.6$ nm. On the other hand, $n_c^{-1/2} \sim 2.7d \sim 1.9$ nm, so our classical theory is correct for $n < n_c$ and needs a revision for larger $n >n_c$ where substantial overlap of wave functions leads to decay of biexcitons into free excitons and eventually to electron-hole plasmas~\cite{Xu2021}. 
This, however, does not change our prediction that at $n < n_c$ biexcitons condense into droplets with $n \sim n_c$, which produces red shifted photoluminescence independent on the light intensity. 

So far we dealt with zero temperature and ignored Coulomb impurities in the system. Our predictions remain quantitatively correct if the temperature $T$ is much smaller than the energy of the crystal per exciton $0.054 e^2/\kappa d \sim 180\,\mathrm{K}$ and concentration of Coulomb impurities $n_i \ll n_c \sim 3 \times 10^{13}$ cm$^{-2}$, where we use $d \sim 0.7$ nm and $\kappa \sim 7$~\cite{Laturia2018}.


\section{Infinite differential capacitance of voltage induced biexciton crystal}
\label{sec:capacitor}
\begin{figure}[t]
	\includegraphics[width=\linewidth]{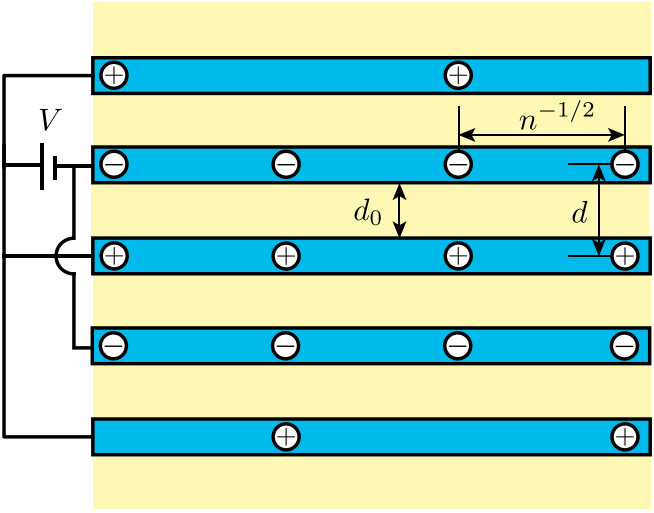}
	\caption
	{Cross section of a capacitor made of five MoSe$_2$ monolayers shown in blue (dark) color.
	hBN spacers of thickness $d_0$ are shown by yellow (light) color. 
	The distance between two charges of an exciton is equal to $d$.
    A voltage $V$ applied between layers 1, 3, 5 and layers 2, 4, induces biexcitons which form a staggered anti-ferroelectric square lattice with a lattice constant $n^{-1/2}$. } \label{fig:capacitor}
\end{figure}

Above we dealt with biexcitons in optically excited WMWMW heterolayers. One can study similar biexciton physics in a device made of five monolayers of MoSe$_2$ separated by identical hBN spacers. If layers 1, 3, and 5 are biased with respect to layers 2 and 4 by a positive voltage $V$,
one can study its nontrivial 135-24 capacitance (c.f. Fig.~\ref{fig:capacitor}). 
Theory of a similar capacitance was the main subject of the previous study of a three-layer device~\cite{Sammon2019}. 
In the case of a pentalayer, all the electrons and holes of neighboring layers form excitons, which in turn form biexcitons with two-dimensional concentration $n$ (c.f. Fig.~\ref{fig:capacitor}). 
Thus, the voltage-biased pentalayer has the same configuration of biexcitons as the contactless but illuminated WMWMW device studied above. 
Therefore, the charge of the 135-24 capacitor per unit area is $2ne$ can be calculated using the energy of the biexciton crystal of a pentalayer $U(n)$ shown in Fig.~\ref{fig:energy}.

\begin{figure}[t]
 	\includegraphics[width=\linewidth]{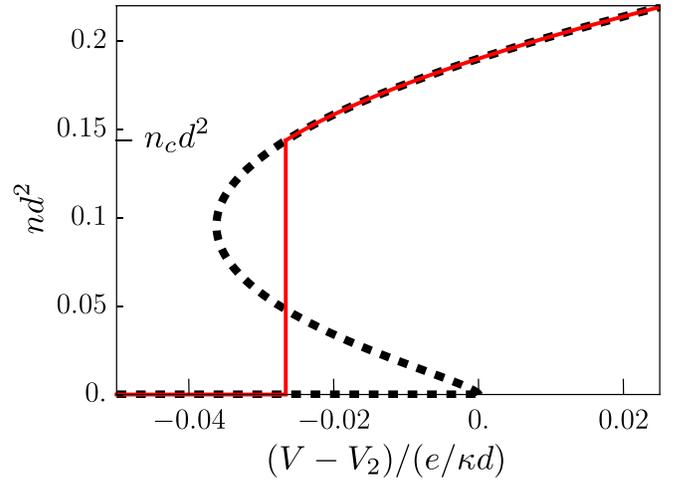}
 	\caption{Dimensionless biexciton density $nd^2$ as a function of dimensionless voltage $(V-V_2)/(e/\kappa d)$ for the pentalayer capacitor shown in Fig.~\ref{fig:capacitor}. The dashed curve shows the function $n(V)$ obtained from Eq.~\eqref{eq:voltage}, while the solid red curve shows the equilibrium function $n(V)$ obtained using the Maxwell's rule Eq.~\eqref{eq:equilibrium}. We see that in equilibrium,  $n(V)$ jumps from zero to $n_c$ at the critical voltage $V_c$.}
 \label{fig:voltage}
\end{figure}

One could expect that the first biexciton enters the capacitor at the voltage
\begin{align}\label{eq:V_2}
    V_2 = (2E_g - E_b )/2e,
\end{align}
where $E_g$ is the band gap of  MoSe$_2$ and $E_b>0$ is the ionization energy of an biexciton (energy necessary to make of two free electrons and two free holes from it). 
Remarkably, at low temperatures, the attraction between biexcitons in the pentalayer device causes a first order phase transition as the applied voltage $V$ grows (c.f. Fig.\,\ref{fig:voltage}). While at small $V$, there are no excitons or biexcitons in the pentalayer and the capacitor remains uncharged, at some critical value $V=V_{c} < V_2$ the whole lattice of alternating biexcitons with concentration $n_c$ emerges. This means that a macroscopic charge $Q_c=2eSn_c$, where $n_c=0.14d^{-2}$ and $S$ is the device area, enters this capacitor. Thus, the differential capacitance $C=dQ/dV$ has a $\delta$-peak at $V = V_{c}$. 

\begin{figure}[t]
 	\includegraphics[width=\linewidth]{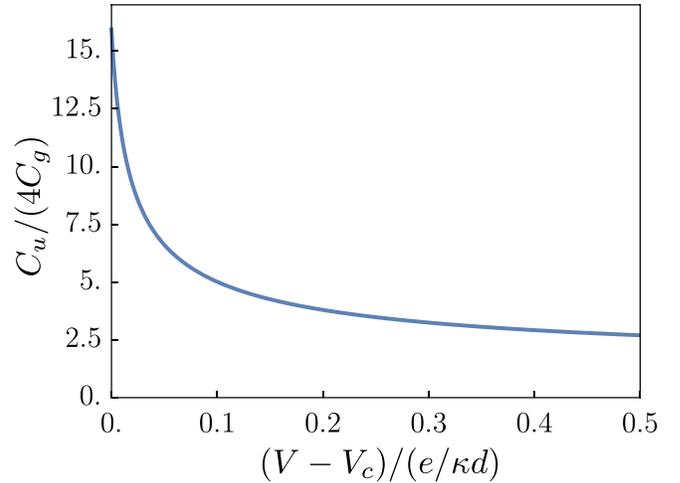}
 	\caption{The non-singular capacitance $C_u/(4C_g)$ as a function of dimensionless voltage $(V-V_c)/(e/\kappa d)$ for the pentalayer MOSe$_2$ device shown in Fig.~\ref{fig:capacitor}. The total geometric capacitance for the pentalayer is equal to $4C_g$, where $C_g = \kappa S/4\pi d$ is the geometric capacitance between two MoSe$_2$ layers.}
 \label{fig:cu}
\end{figure}

The differential capacitance of such a device can be determined from the total electrostatic energy $E$ of the system as 
\begin{align}\label{eq:Cap_def}
    C^{-1}=\frac{d^2E}{dQ^2}=\frac{1}{(2eS)^2}\frac{d^2E}{dn^2}.
\end{align}
The energy $E$ of this system of classical charges can be written as 
\begin{align}\label{eq:Energy_tot}
    E = 2enSV_2 + nSU,
\end{align}
where $V_2$ is the voltage necessary to create a single isolated biexciton and is given by Eq.~(\ref{eq:V_2}), while $U$ is the interaction energy of the anti-ferroelectric biexciton crystal per biexciton. 
Using Eqs.~\eqref{eq:U_total} and \eqref{eq:Energy_tot}, one can calculate the voltage as
\begin{align}\label{eq:voltage}
    V = \frac{dE}{dQ} = V_2 + \frac{1}{2e} \frac{d(nU)}{dn}.
\end{align}
Eq.~\eqref{eq:voltage} results in the dimensional density $n(V)d^2$ as a function of the voltage $(V-V_2)/(e/\kappa d)$ shown by the dashed curve in Fig.~\ref{fig:voltage}. 
Note that in a range of voltages this curve has three branches: a lower branch $n=0$, a middle branch, and an upper branch. 
Within the middle branch, the differential capacitance defined by Eq.~\eqref{eq:Cap_def} is negative, so that this region is thermodynamically unstable and inaccessible. 
Thus, in experiments, we do not expect the density to change continuously along the dashed curve, but instead it should follow the red full curve where the density of biexcitons $n$ jumps from zero to a finite value $n_c=0.14d^{-2}$ at the critical voltage 
\begin{equation}\label{eq:V_c}
    V_c=V_2-0.027\frac{e}{\kappa d}.
\end{equation} 
Here $V_c$ is determined by the Maxwell area rule~\cite{landau_statphys1}
\begin{equation}\label{eq:equilibrium}
    \int_0^{n_c} n(V)dV=0,
\end{equation}
where the integral is taken along the dashed curve in Fig.~\ref{fig:voltage}. At $V=V_c$ the two regions lying between the vertical red line and the dashed curve have equal areas.
It is worth noting that $n_c$ obtained from Maxwell's area rule is the same $n_c$ at which $U$ reaches its minimum value (c.f. Fig.~\ref{fig:energy}). 
As the density abruptly jumps, there is a $\delta$-peak in the capacitance at $V=V_c$. For $V\geq V_c$ we can write the capacitance as
 \begin{equation}\label{eq:C(V)}
 C(V) = 2eSn_c\delta(V-V_c) + C_u(V),
 \end{equation} 
 where the non-singular capacitance $C_u(V) = 2eS(dn/dV)$ is obtained by differentiating the upper branch of the $n(V)$ red curve shown in Fig.~\ref{fig:voltage} with respect to $V$, and the result is shown in Fig.~\ref{fig:cu}. 
 As $V$ approaches $V_c$ from above, $C_u(V)$ grows as $(V-V_c+0.01e/\kappa d)^{-1/2}$, and reaches a very large maximum value $C_u(V_c)\simeq 64 C_g$, where $C_g = \kappa S/4\pi d$ is the geometrical capacitance of the capacitor formed by two MoSe$_2$ planes separated by distance $d$. (The origin of small number 0.01 and large number 64 can be traced to small value $n_{c}d^2=0.14$).
 
Above we dealt with the strong effect of correlations between electrons and holes, which makes the capacitance much larger than the geometrical one. Historically, relatively small addition to the geometrical capacitance of a plane capacitor due 
to strong electron correlations leading to the negative compressibility of electron gas in one of the capacitor plates was predicted long ago~\cite{Bello1981} 
and observed in Si MOSFET~\cite{Kravchenko1990} and GaAs two-dimensional gas~\cite{Eisenstein1992}. 
Later it was predicted that in a thin capacitor with a small charge, electron and holes form compact dipoles weakly repelling each other at large distances. 
As a result, its capacitance becomes much larger than the geometrical one~\cite{Skinner2010}. 
Indeed, a substantial capacitance enhancement was simultaneously observed~\cite{Li2011}.

In this section, we predicted that in the clean pentalayer capacitor, the 135-24 differential capacitance diverges at some critical voltage at zero temperature due to attraction of biexcitons. 
Such a divergence was previously predicted for two other kinds of capacitors, where attraction between dipoles dominates their repulsion~\cite{Chen2011,Sammon2019}. 
Of course, in the real world the maximum capacitance has a much larger than geometrical but finite value, due to a finite disorder and a finite temperature.

\section{Four-layer devices}
\label{sec:fourlayer}
\begin{figure}[t]
	\includegraphics[width=\linewidth]{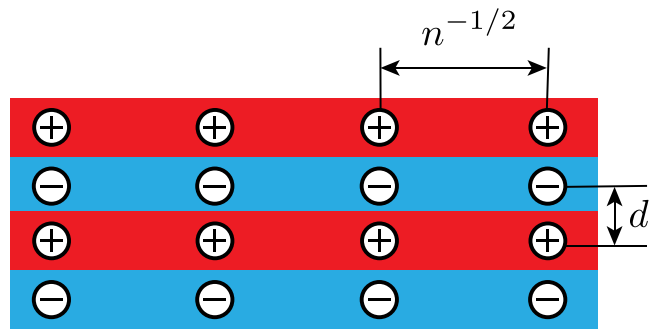}
	\caption{Asymmetric four layer WSe$_2$/MoSe$_2$/WSe$_2$/MoSe$_2$ device. WSe$_2$ layers are red and MoSe$_2$ layers are blue. When the four layer device is illuminated at low temperatures, identical linear biexcitons which repel each other are formed.}\label{fig:tetralayer}
\end{figure}

So far in this paper we dealt with pentalayer devices and their comparison with trilayer ones. 
Let us briefly dwell on the contactless, but illuminated by interband light, four layer devices.
In a asymmetric four layer WMWM device, all excited by light excitons at small density bind into identical linear dipolar biexcitons, which repel each other and form a ferroelectric (see Fig.~\ref{fig:tetralayer}). 

Alternatively, one can imagine a symmetric four layer WMMW device, where oppositely directed excitons emerge in the two bottom and the two upper layers. They attract each other at large distances and repel at small ones. The ground state of such four layer device is an anti-ferroelectric staggered crystal similar to the one shown in Fig.~\ref{fig:trilayer} for WMW layer. However, for a given two-dimensional concentration of excitons $n$, both attraction and repulsion are weaker. We found that the staggered crystal cohesive energy per exciton in a WMMW tetralayer is 8 times smaller than in a WMW trilayer. A single exciton in the upper two layers can in principle hybridize with the one in the lower two layers, resulting in a quadrupolar dark exciton. But the related matrix element and the quantum energy splitting are probably much weaker even than the Coulomb energy cost of inverting an exciton dipole moment in a staggered crystal and therefore can be ignored. 

\begin{acknowledgments}
We are grateful to L. Butov, M.M. Glazov, M. Sammon and X.-Y. Zhu for helpful discussion. 
Y.H. gratefully acknowledges support from Larkin Fellowship at the University of Minnesota. 
\end{acknowledgments}


%

\end{document}